# Giant Splitting of Folded Dirac Bands in Kekulé-ordered Graphene with Eu Intercalation

Xiaodong Qiu,[1]† Tongshuai Zhu,[2]† Zhenjie Fan,[1] Kaili Wang,[1] Yuyang Mu,[1] Bin Yang,[1] Di Wu,[1,3,4] Haijun Zhang,[1,3,4] Can Wang,[5,3]* Huaiqiang Wang,[6]* Yi Zhang[1,3,4,7]*

[1] *National Laboratory of Solid State Microstructure, School of Physics, Nanjing University, Nanjing 210093, China.*

[2] *Collage of Science, China University of Petroleum (East China), Qingdao 266580, China.*

[3] *Collaborative Innovation Center of Advanced Microstructures, Nanjing University, Nanjing 210093, China.*

[4] *Jiangsu Physical Science Research Center, Nanjing University, Nanjing 210093, China.*

[5] *Hunan Provincial Key Laboratory of Flexible Electronic Materials Genome Engineering, School of Physics and Electronic Science, Changsha University of Science and Technology, Changsha 410114, China.*

[6] *Center for Quantum Transport and Thermal Energy Science, School of Physics and Technology, Nanjing Normal University, Nanjing 210023, China.*

[7] *Hefei National Laboratory, Hefei 230088, China.*

* Corresponding authors' Email:  jerrywang@csust.edu.cn
hqwang@njnu.edu.cn
yizhang@nju.edu.cn

† These authors contributed equally to this work.

**Abstract**

Kekulé-ordered graphene on SiC realized by intercalating two-dimensional metal layers offers a versatile platform for exploring intriguing quantum states and phenomena. Here, we achieve the intercalation of $(\sqrt{3} \times \sqrt{3})R30°$-ordered Eu layer between epitaxial graphene and SiC substrate, realizing a Kekulé graphene with large local magnetic moments of intercalated Eu atoms. Combining angle-resolved photoemission spectroscopy (ARPES) and density functional theory (DFT) calculations, we revealed that the Kekulé order folds the Dirac cones of graphene from the corners to the Brillouin zone center via intervalley scattering, forming the replica Dirac bands with gap opening. More intriguingly, the Dirac fermions in the replica Dirac bands show a strong exchange coupling with the localized magnetic moments of Eu $4f$ orbitals, resulting in a giant splitting of the folded Dirac bands. The observation of strong coupling between Dirac fermions and local magnetic moments of Eu $4f$ electrons via Kekulé order pave a new way for generating Dirac band splitting in graphene, advancing the potential applications of Kekulé-ordered graphene in spintronics, as well as exploring intriguing physical properties and correlation states for quantum technology.

**Introduction**

Many-body interaction within two-dimensional (2D) limits is a hot topic in condensed matter physics. In particular, interacting Dirac electrons in flat bands with strong correlation give rise to a variety of quantum phases [1]. Graphene with a $(\sqrt{3} \times \sqrt{3})R30°$ Kekulé distortion is a prominent platform for exploring various novel quantum phenomena, such as electron fractionalization [2], chiral symmetry breaking and gap opening [3, 4], unconventional superconductivity [5, 6], valley-polarization and valleytronics [7-11]. The Kekulé order can be realized by the interference of Dirac fermions scattered by impurities on graphene, which were directly revealed by scanning tunneling microscopy (STM) [12-16], and the folded replica Dirac bands at Brillouin zone (BZ) center were also directly observed by angle-resolved photoemission spectroscopy (ARPES) [17, 18]. Another promising progress for realizing Kekulé order in graphene is element intercalation. Especially, Kekulé-ordered chiral symmetry breaking and gap opening have been verified in Li- intercalated graphene on SiC [4]. In addition to the Kekulé order, element intercalation in graphene presents a promising approach for exploring the interaction between Dirac fermions and additional novel electronic states. In particular, the intercalation of lanthanide elements (such as Ce, Eu, Gd and Tb), which exhibit large localized magnetic moments of $4f$ orbitals, is essential of disentangling interplay between Dirac fermions and heavy quasiparticles within a unique 2D platform. For examples, Ce-intercalated graphene exhibits Kondo resonance, highlighting the interplay between Dirac fermions and strongly correlated electron systems [19]. The intercalation of Eu and Gd in graphene induces spin-dependent effects and Lifshitz transitions [20, 21], while Tb intercalation leads to an topological superconductivity state [22], collectively offering new avenues for tunable electronic and magnetic properties in 2D materials. The intercalation of graphene with various elements enables the study of many-body interactions, providing a platform for exploring emergent quantum phenomena.

Since the half-filled $4f$ orbitals of Eu possesses the largest magnetic moment among the lanthanide elements, the Eu interaction in graphene has attracted extensive research interest recently. The Eu- intercalated graphene on Ir(111) can form either $(2 \times 2)$ or $(\sqrt{3} \times \sqrt{3})R30°$ superlattice, which exhibit different magnetic behaviors [23]. Previous studies have addressed the hybridization of the graphene π-band with the Eu $4f$ band in Eu-intercalated graphene on SiC [24, 25]. Recently, the intercalation of Eu underneath graphene on Co(0001) allowed single flat band formation at the Fermi level due to hybridization between graphene, Eu and Co [20]. Eu-intercalated graphene on Ni(111) with a $(\sqrt{3} \times \sqrt{3})R30°$ superlattice shows a spin-dependent gap observed by spin-resolved ARPES [26]. Despite previous studies reporting the presence of a $(\sqrt{3} \times \sqrt{3})R30°$ superlattice in Eu-intercalated graphene on Ir(111), Ni(111), Co(0001) or SiC, the expected feature of Kekulé phase with characterized replica Dirac band at the BZ center is still not unveiled. More important, the Dirac fermions of the replica Dirac band, originated from the scattering with Kekulé order formed by Eu intercalation, would show strong coupling with the localized magnetic moments of $4f$ electrons, which could generated a vast variety of exotic quantum phases and phenomena [27].

In this work, we realized the $(\sqrt{3} \times \sqrt{3})R30°$ Kekulé ordered intercalation of Eu underneath the epitaxial monolayer graphene (MLG) on SiC(0001) substrate by using molecular beam epitaxy (MBE). In addition to the heavily electron doping on the MLG, the replica Dirac bands folded at the BZ center was clearly observed by the *in-situ* APRES. More interestingly, the folded Dirac bands show a giant Zeeman-type splitting in energy, resulting a complex Fermi surface pattern. While the intrinsic Dirac cone of MLG at the BZ corners keeps its dispersion feature without spitting. Combining the density functional theory (DFT) calculations, we reveal that the folding process of the Dirac bands, originated from intervalley scattering by means of the Kekulé order, show a strong exchange coupling with the localized magnetic moments of the Kekulé-ordered Eu $4f$ orbitals. The calculated Dirac band splitting and Fermi surface pattern based on this coupling mechanism agree well with the experimental ARPES results. The intrinsic Dirac band of MLG, since not participating in the band folding process, shows no splitting feature. Besides, the replica Dirac bands show a strong hybridization with the Eu $4f$ bands at the BZ center. The observation of strong coupling between Dirac fermions and local magnetic moment of Eu $4f$ electrons via Kekulé order pave a new way for generating Dirac band spitting in graphene, advancing the potential applications of Kekulé-ordered graphene in spintronics, as well as exploring intriguing physical properties and correlation states for quantum technology.

## Results

The MLG was synthesized by thermal decomposition of SiC, with a C buffer layer formed underneath the MLG, as shown in Fig. 1(a). A well intercalated Eu layer is located between the MLG and buffer layer, forming the $(\sqrt{3} \times \sqrt{3})R30°$ order, as shown in Fig. 1(b)&(c). In the x-ray photoemission spectroscopy (XPS) results shown in the Supplementary Materials part A, the emerging Eu $3d$ peaks and the no shift of the Si $2p$ peak can evidence that the Eu atoms are intercalated between the buffer layer and MLG (24). Fig. 1(d-f) are the RHEED patterns of MLG, partially Eu-intercalated MLG and fully Eu-intercalated MLG (Eu-MLG). The MLG-$(1 \times 1)$ diffraction spots became weaker during the intercalation process, and were completely replaced by the Eu-$(\sqrt{3} \times \sqrt{3})R30°$ diffraction spots after fully Eu-intercalation. Fig. 1(g) presents the STM image on a partially Eu-intercalated MLG. The Eu-MLG region shows a higher height of ~3.5 Å than the MLG region, indicating the existence of an intercalated Eu layer. Fig. 1(h)&(i) are the atomically resolved STM images taken on the surfaces of MLG and Eu-MLG, respectively. The corresponding fast Fourier transform (FFT) image of MLG only shows the $(1 \times 1)$ spots (indicated by the blue circles), while the FFT image of Eu-MLG shows the Eu-$(\sqrt{3} \times \sqrt{3})R30°$ spots (green circles) in addition to the $(1 \times 1)$ spots. The evolution of the RHEED patterns and the STM images clearly evidence the well-ordered intercalation of the Eu atoms as a $(\sqrt{3} \times \sqrt{3})R30°$ superlattice, forming the Kekulé-ordered graphene.

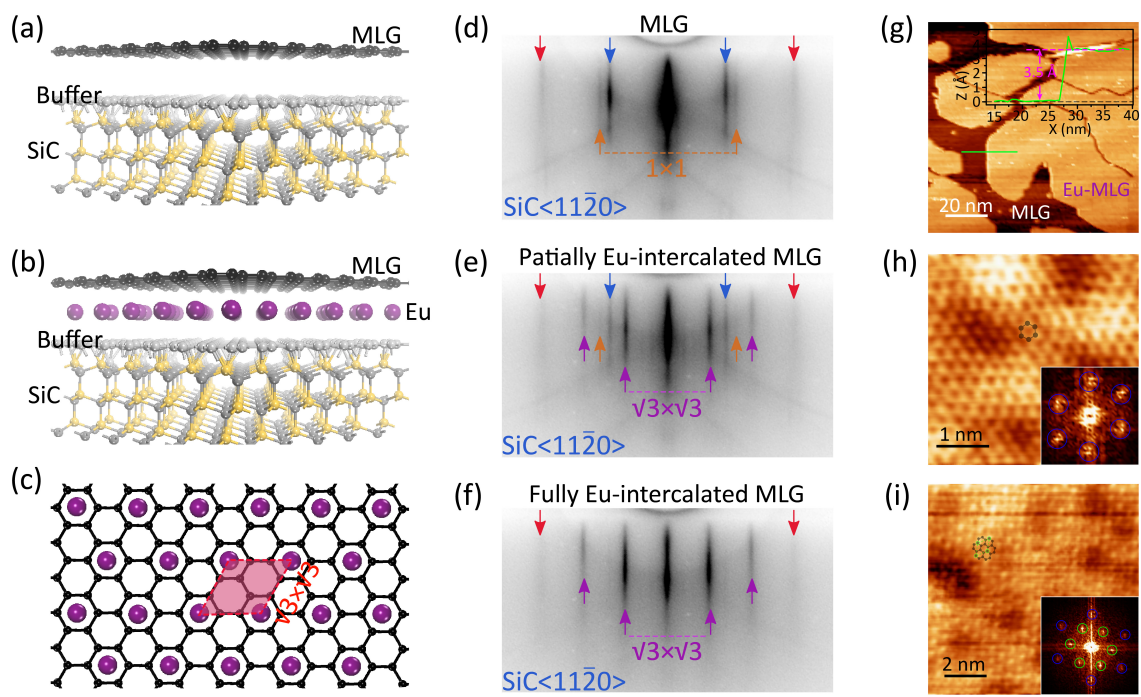

**Fig. 1. Growth of Eu intercalated MLG on SiC. (a-c)** Schematic illustration of the intercalation process of Eu into MLG. **(d-f)** Evolution of the RHEED patterns along the direction of SiC< $11\bar{2}0$ > during Eu intercalation process. The RHEED patterns highlight the presence of SiC-$(1 \times 1)$, MLG-$(1 \times 1)$ and intercalated Eu-$(\sqrt{3} \times \sqrt{3})R30°$ diffraction spots, indicated by the blue, yellow, and purple arrows, respectively. **(g)** STM image of a partially Eu-intercalated MLG with the inset height profile along the green line. **(h-i)** Atomically resolved STM images of (h) MLG (4×4 nm²) and (i) Eu-MLG (10×10 nm²) taken at room temperature (300 K), respectively. Insets are the FFT images of them, the blue and green circles indicate the MLG-$(1 \times 1)$ and intercalated Eu-$(\sqrt{3} \times \sqrt{3})R30°$ spots, respectively.

To investigate the electron structures modulated by the Kekulé order of Eu intercalation, we have performed ARPES measurements on the MLG and Eu-MLG in Fig. 2. Before the Eu intercalation, the characterized Dirac cones of MLG located at the K points of BZ can be observed in Fig. 2(a)&(b), with a small band gap opening at the Dirac point due to graphene-substrate interaction (28). After intercalation, the Kekulé order results in a folded BZ to the original one of MLG [indicated by the dashed green and solid red hexagons in Fig. 2(c)&(d)]. The Fermi surface mapping shown in Fig. 2(c) clearly shows the enlarged Dirac pockets around the K/K' points of MLG BZ, indicating the heavy electron doping effect from the intercalated Eu to MLG. Notably, by enhancing the intensity of Fermi surface mapping in Fig. 2(d), a complex Fermi surface pattern can be distinguished around the Γ point. A zoom-in mapping is shown in Fig. 2(e) for a clear presentation of this pattern. Many investigations have revealed that the Kekulé-ordered graphene can result in the folding of Dirac bands from the K/K' points into the Γ point, forming a chiral-symmetry-breaking hexagram-like Fermi pocket due to the overlapping of chiral K/K' Dirac cones (4, 17, 18). However, in Fig. 2(e), the Fermi surface shows a feature consisting of two sets of hexagram-like pockets depicted by the red and green dashed lines, which implies the splitting of the folded Dirac bands. This band splitting can be further characterized by the ARPES spectra along the cut #1 and cut #2 directions shown in Fig. 2(f)&(g). Along the cut #1 direction, the splitting of folded Dirac bands shows two bands crossing the Fermi level, indicated by the red and green arrows and dashed lines shown in Fig. 2(f). Along the cut #2 direction, the splitting Dirac bands show four bands crossing the Fermi level, while the middle two branches merge into almost one crossing point at the Fermi level. The localized Eu $4f$ orbitals show the characterized flat bands at $-1.5 \sim -2.0$ eV in Fig. 2(f-h) (20, 26).

In Fig. 2(c)&(d), we note that the ARPES signal around the K/K' points is much stronger than other regions. The ARPES spectra shown in Fig. 2(i)&(j) reveal the heavily electron-doped Dirac cone with an enlarged band gap (0.63±0.02 eV) at the Dirac point, which agrees with the previous ARPES results on Eu intercalated graphene (20, 26). The folded Dirac bands around the Γ point show distinctly weaker ARPES intensity than the Dirac cones around the K/K' points. Therefore, we suggest that only minor Dirac fermions of MLG participate in the band folding process of Kekulé order, in contrast, the major Dirac fermions of MLG result in the much stronger signal at the K/K' points. Meanwhile, only the folded Dirac bands show the splitting feature, while the intrinsic Dirac bands without folding show no splitting.

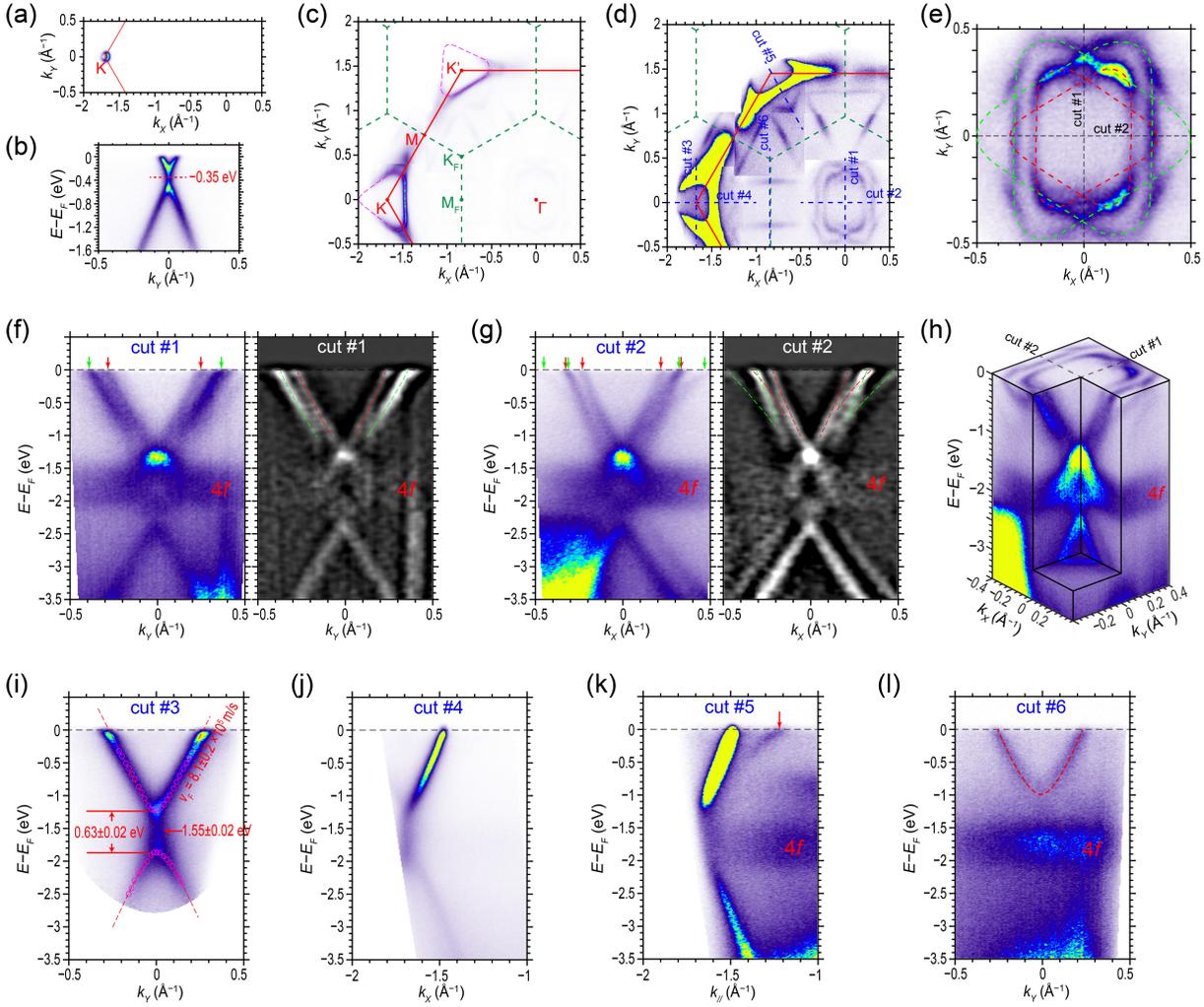

**Fig. 2. ARPES spectra of Eu-MLG. (a-b)** ARPES Fermi surface mapping (a) and spectra cut (b) of MLG, respectively. **(c)** ARPES Fermi surface mapping of Eu-MLG. **(d)** Enhanced intensity of (c). The solid red and dashed green hexagons are the BZs of MLG and $\left(\sqrt{3} \times \sqrt{3}\right)R30°$ Kekulé superlattice, respectively. **(e)** Zoom-in Fermi surface mapping of Eu-MLG around the $\Gamma$ point. **(f-g)** ARPES spectra and corresponding second-derivatives of Eu-MLG along the (f) cut #1 and (g) cut #2 directions, respectively. **(h)** 3D view of the ARPES spectra of Eu-MLG around the $\Gamma$ point. **(i-l)** APRES spectra of Eu-MLG along the (i) cut #3, (j) cut #4, (k) cut #5 and (l) cut #6, respectively.

The $\Gamma$ point of the second folded BZ of the Kekulé superlattice overlaps with the K/K' points of MLG BZ, as shown in Fig. 2(c)&(d). Therefore, the folded Dirac bands with splitting should also appear around the K/K' points. However, due to the rather strong signal from the intrinsic Dirac band and matrix element effects in ARPES (*29, 30*), it is difficult to distinguish the splitting of folded Dirac bands at the K point. But around the K' point in Fig. 2(d), a weak signal from the folded Dirac bands can be observed in addition to the intrinsic Dirac pocket along the cut #5 direction, indicated by the red arrow in Fig. 2(k).

In addition to the folded and intrinsic Dirac bands around the Γ and K/K' points in Fig. 2(d), additional triangular Fermi pockets can be observed around the $K_F$ points. In Fig. 2(i), this pocket shows a parabolic dispersion band at the $M_F$ point depicted by the red dashed line, and will be verified to be the hybridized C-2$p$ and Eu-4$f$ orbitals of $(\sqrt{3} \times \sqrt{3})R30°$ Eu-MLG superlattice (see Supplementary Materials part B).

To elucidate the physical origin of the observed band splitting in its folded Dirac cones, we performed first principal calculations and tight-binding modeling of the Kekulé-ordered Eu-MLG heterostructure. Prior to presenting quantitative results, we provide a qualitative description of the three-stage formation process of the spin-split Dirac bands, as shown in Fig. 3(c). First, in the $(\sqrt{3} \times \sqrt{3})R30°$ superlattice BZ, the Dirac cones originated from the K and K' valleys of MLG fold back to the Γ point, resulting in gapless replica Dirac bands at Γ. Second, the emergence of the Kekulé-O type bond texture featuring two distinct hopping strengths $t$ and $t'$ between nearest-neighbor carbon atoms [see Fig. 3(a)], induces intervalley couplings between the two replica Dirac cones and opens an energy gap. Third, the strong exchange coupling $J \cdot \mathbf{S}_{Eu} \cdot \mathbf{s}$ between local Eu moments ($\mathbf{S}_{Eu}$) and Dirac electron spins ($\mathbf{s}$), as shown in Fig. 3(b), lifts the spin degeneracy of the gaped Dirac cones, leading to an energy splitting of $\Delta$.

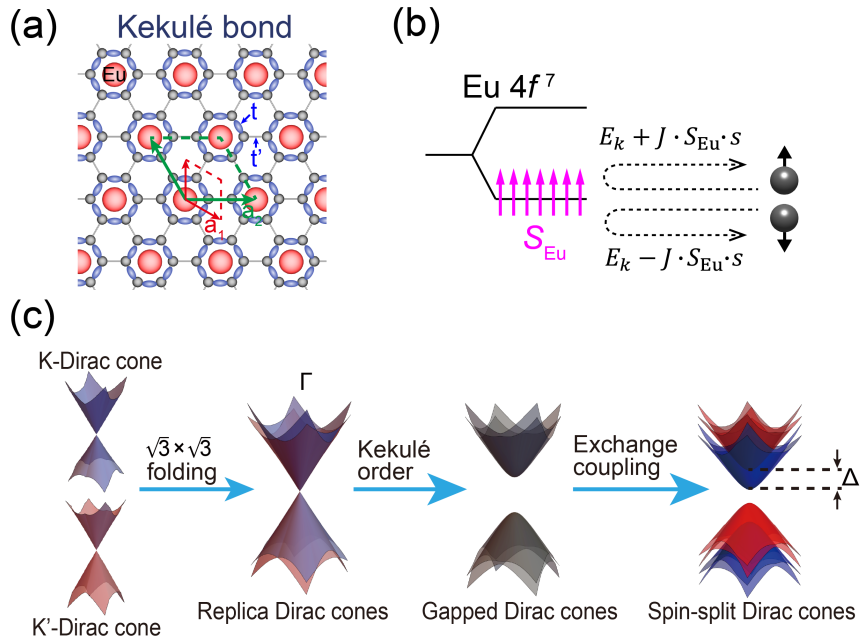

**Fig. 3. Formation of the spin-split Dirac bands in Eu-MLG. (a)** Schematic of the Kekulé-O type bond ordering in Eu-MLG, with distinct hopping strength $t$ and $t'$ between nearest-neighbor carbon atoms. **(b)** Illustration of the exchange coupling between Dirac electron spins and local magnetic moments originating from half-filled Eu-4$f$ orbitals. **(c)** Schematic of the three-stage process for the formation of spin-split Dirac cones at the Γ point in folded BZ, sequentially considering the $\sqrt{3} \times \sqrt{3}$ superlattice BZ folding, Kekulé bond ordering, and exchange coupling.

Figure 4(a) shows the DFT orbital-projected band structure near the Fermi level of the $(\sqrt{3} \times \sqrt{3})R30°$ superlattice BZ, where Eu-4$f$ orbitals from narrow bands are around the Dirac point, which is consistent with ARPES results. The hybridized C-2$p$ and Eu-4$f$ orbitals form the electronic pockets around the $K_F$ points of folded BZ and the dispersing band along cut #6 crossing Fermi level (see the Supplementary Materials part B). Consequently, the Eu-4$f$ bands are expected to have very strong hybridization and coupling with the folded Dirac bands at the $\Gamma$ point. Note that since almost no long-range magnetic order is observed experimentally (see the Supplementary Materials part C), we employed spin-unpolarized DFT calculations, and thus Dirac band splitting is absent in Fig. 4(a). Based on the three-stage process mentioned above, further including the two distinct hopping strengths and the exchange coupling with local Eu moments, the gap opening and band splitting will emerge in the folded Dirac bands (see the Supplementary Materials part D). By extracting the moment distribution curves (MDCs) and energy distribution curves (EDCs) from the ARPES results shown in Figs. 4(e)&(g), we can experimentally determine the energy splitting size of $\Delta = 0.47 \pm 0.02$ eV, and the moment splitting sizes along the cut #1 and cut #2 directions at Fermi level are $0.116 \pm 0.005$ Å$^{-1}$ and $0.097 \pm 0.005$ Å$^{-1}$, respectively. To account for the exchange coupling with local Eu moments, we construct a tight-binding model of the Kekulé-ordered MLG, incorporating distinct nearest-neighbor hopping strength $t$ and $t'$, along with a 0.235 eV exchange filed (Zeeman-type term) obtained from experimental data $\Delta/2$. Figures 4(c)&(d) compare Fermi contours around the $\Gamma$ point from ARPES measurements and tight-binding calculations at four energies, namely, 0 eV, −0.3 eV, −0.6 eV, and −0.9 eV. Close agreement is observed across these levels, with particularly strong consistency at the Fermi level (0 eV), as further evidenced in Fig. 4(b) exhibiting spin-split Fermi contours. In addition, tight-binding dispersions for the spin-split folded Dirac bands along cut #1 and cut #2 are plotted in Figs. 4(f)&(h) as the red curves, respectively, demonstrating good agreement with corresponding ARPES results in Figs. 4(e)&(g). The strong intensity appeared at −1.4 eV in the ARPES spectra Figs. 4(e)&(g) could be attributed to the hybridization effect between the Eu $4f$ bands and spitting folded Dirac bands at the $\Gamma$ point, which is difficult to directly simulate by the DFT calculations.

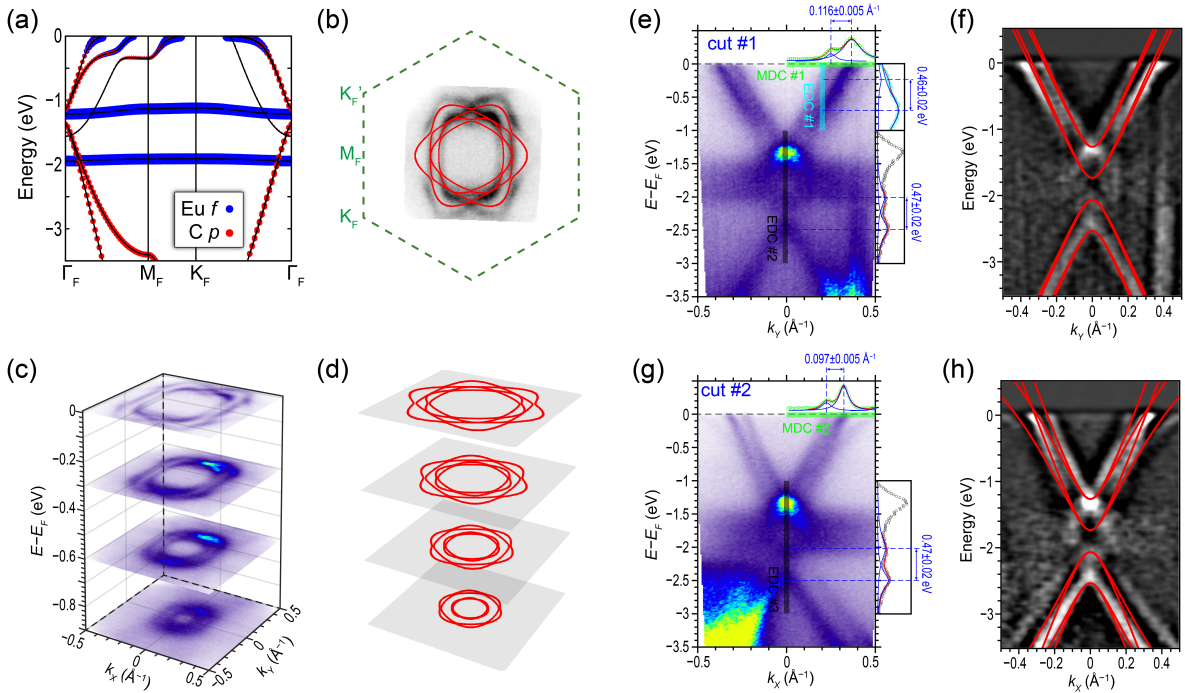

**Fig. 4. Comparison between the calculated and experimental results. (a)** DFT orbital-projected band structure of the Eu-MLG superlattice near the Fermi level, where blue and red denote Eu-4$f$ and C-2$p$ orbitals, respectively. **(b)** Superposed plot of the calculated Fermi contour overlapping the ARPES data near the $\Gamma$ point. **(c-d)** Fermi contours at the four energies from the ARPES data (c) and tight-binding calculations (d). **(e-h)** ARPES spectra with the corresponding MDCs and EDCs (e, g) along the cut #1 (e, f) and cut #2 (g, h) directions, and the tight-binding simulations overlapping the second-derivative ARPES spectra (f, h), demonstrating close agreement with each other.

**Discussion**

In summary, we synthesize the Kekulé-ordered graphene by periodically intercalating Eu atoms into the epitaxial MLG on SiC substrate. In this system, the Kekulé order induced intervalley coupling results in the folded Dirac bands at the $\Gamma$ point. More intriguingly, this intervalley coupling process further simultaneously involve a strong exchange coupling effect with the local moments of intercalated Eu atoms, driving the giant splitting of the folded Dirac bands. The strong exchange coupling effect with local moments provide a new modulation freedom of the Dirac fermion, in addition to the band folding effect via Kekulé order. The giant band splitting due to exchange coupling could expand the potential applications of Kekulé-ordered graphene in spintronics, as well as exploring intriguing physical properties and correlation states for quantum technology.

**Materials and Methods**

- **Sample preparation**

The MLG and Eu-MLG samples were prepared in a combined MBE-STM-ARPES ultrahigh vacuum (UHV) system with a base pressure of $1.5 \times 10^{-10}$ mbar. The MLG was synthesized by flashing the 4H-SiC(0001) wafer at 1450 ℃ for 2 cycles, and each cycle was kept for 15 seconds. The high purity Eu was evaporated from a standard Knudsen cell at 430 ℃. During the intercalation process, the MLG was kept at 300 ℃. A fully Eu-intercalated MLG was obtained by evaporating Eu on MLG for 13 minutes. Shorter or longer time of intercalation will result in a partially Eu-intercalated MLG or intercalation of Eu underneath the buffer layer. The synthesis process of MLG and Eu-MLG were monitored by an *in-situ* RHEED system.

- **Experimental characterizations**

The sample surface morphology was characterized by an *in-situ* STM working at room-temperature (300 K). The *in-situ* ARPES and XPS measurements were performed via a shared Scienta Omicron DA30 analyzer. The monochromatic ultraviolet (UV) light for ARPES was generated from a helium lamp (He I mode, $h\nu$ = 21.218 eV). The monochromatic x-ray for XPS was generated from an Al electrode excitation source (Al Kα, 1486.7 eV). During the ARPES and XPS measurements, the samples were cooled down to the lowest ~10 K by a closed-cycle cryo-generator.

- **Theoretical calculations**

We performed the first-principles calculations by employing the Vienna ab-initio simulation package (VASP) *(31, 32)*. The calculations employed the generalized gradient approximation (GGA) with the Perdew-Burke-Ernzerhof (PBE) *(33)* type exchange-correlation potential is adopted, setting the energy cutoff at 600 eV. A GGA+U functional with the Hubbard $U$ parameter set at $U$ = 6 eV for Eu-4$f$ orbitals is considered in this work. To obtain the lattice structure and atomic positions of Eu-intercalated MLG, we adopted a vacuum layer of 15 Å, and performed relaxation with fixed unit cell volume until the force less than $1.0 \times 10^{-3}$ eV/Å. The superlattice parameter was determined as 4.29 Å, with a distance of 2.35 Å between Eu and MLG. To properly account for the van der Waals (vdW) interactions, the structural relaxation was performed employing the vdW-DF2 *(34)* with an optB88-vdW functional *(35, 36)*. The k-point sampling grid of the Brillouin zone in the self-consistent process was performed using a -centered Monkhorst-Pack k-point mesh of 8×8×1. A total energy tolerance $10^{-8}$ eV was adopted for self-consistent convergence.

In the tight-binding simulations of the Kekulé-ordered Eu-MLG superlattice, the two distinct nearest-neighbor hoppings between carbon atoms are chosen as $t$ = −2.8 eV and $t'$ = −3.2 eV, respectively. The Zeeman-type exchange coupling is chosen as 0.235 eV to fit with the experimentally observed 0.47 eV splitting. The Fermi level is set at 2 eV in the tight-binding simulations to fit with experiments.

**Acknowledgments**

**Funding:**

National Natural Science Foundation of China 92165205

Innovation Program for Quantum Science and Technology 2021ZD0302803

Fundamental Research Funds for Central Universities (Nos. 0204/14380228, 23CX06063A)

Fundamental Research Program of Natural Science Foundation of Jiangsu Province BK20243011

Natural Science Foundation of Jiangsu Province BK20233001

**Author contributions:**

Conceptualization: YZ, XQ, CW
Methodology: YZ, XQ, CW, HW, DW
Investigation: XQ, TZ, ZF, KW, YM, BY
Visualization: XQ, TZ, YZ
Supervision: YZ, HW, HZ, DW
Writing—original draft: CW, XQ, TZ
Writing—review & editing: YZ

**Competing interests:**

Authors declare that they have no competing interests.

**Data and materials availability:**

All data are available in the main text or the supplementary materials.

## Supplementary Materials

The Supplementary Materials contains:

A. XPS of the MLG and Eu-MLG;

B. Band structure of the $\left(\sqrt{3} \times \sqrt{3}\right)R30°$ Eu-MLG superlattice;

C. Non-long-range magnetic order of Eu-MLG;

D. Tight-binding calculation based on the three-stage process

# Supplementary Materials for

## Giant Splitting of Folded Dirac Bands in Kekulé-ordered Graphene with Eu Intercalation

Xiaodong Qiu *et al.*

*Corresponding author. Email: zhangyi@nju.edu.cn

**This PDF file includes:**

Supplementary Text:

A. XPS of the MLG and Eu-MLG
B. Band structure of the $\left(\sqrt{3} \times \sqrt{3}\right)R30°$ Eu-MLG superlattice
C. Non-long-range magnetic order of Eu-MLG
D. Tight-binding calculation based on the three-stage process

Figs. S1 to S4

**Supplementary Text**

### A. XPS of the MLG and Eu-MLG

Figures S1(a)&(c) show the XPS results on the Si 2*s* and 2*p* core levels of MLG and Eu-MLG samples, respectively. In Fig. S1(a), the core levels of Si 2*s* and 2*p* are located at 152.4 eV and 101.2 eV, respectively. After the fully intercalation of Eu, the core levels of Si 2*s* anf 2*p* show no shift, while the peaks of Eu 4*d* emerges between 134.6 eV and 128.8 eV [Fig. S1(c)]. According to the previous report, the no shift of Si 2*s* and 2*p* core levels indicate the intercalated Eu atoms are located bettwen the buffer layer and MLG, but not underneath the buffer layer *(24)*.

Figure S1(b) shows the C 1*s* core levels of the MLG sample, which is constitued of three peaks from the SiC substrate (283.3 eV), the buffer layer (284.9 eV), and the graphene layer (Gr, 284.0 eV). After the fully intercalation of Eu, the core levels of the buffer laeyr and Gr shift to higher binding energies (286.2 eV and 285.3 eV) by approximately 1.3 eV due to the electronic doping from Eu, consisting with the ARPES results shown in Fig. 2. The Gr peak also becomes stronger after the Eu intercalation.

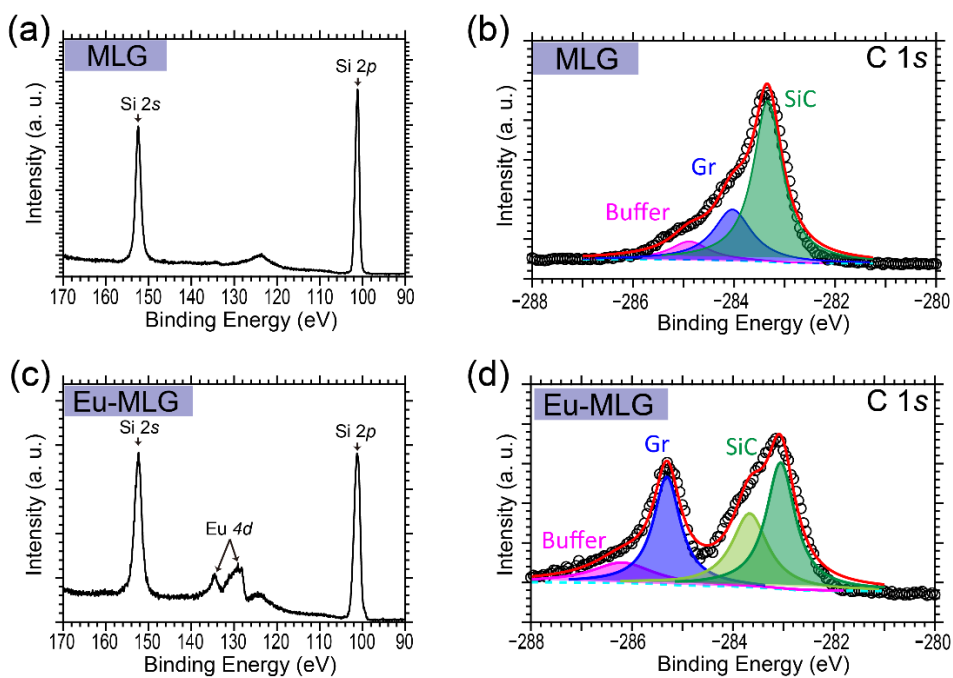

**Fig. S1. XPS spectra of the MLG and Eu-MLG. (a)&(b)** XPS results of the MLG around the Si 2*s* and 2*p* core levels (a), and the C 1*s* core levels (b). **(c)&(d)** XPS results of the Eu-MLG around the Si 2*s* and 2*p* core levels (c), and the C 1*s* core levels.

## B. Band structure of the $(\sqrt{3} \times \sqrt{3})R30°$ Eu-MLG superlattice

Figure S2 show the calculated band structure of the $(\sqrt{3} \times \sqrt{3})R30°$ Eu-MLG superlattice overlapping the corresponding ARPES spectra. The hybridized Eu-4$f$ and C-2$p$ orbitals ($f$-$p$) forms an electronic pocket around the $K_F$ point of the folded BZ in Fig. S2(a). In Fig. S2(b), the 4$f$ bands of Eu superlattice lay between −1.1 eV and −1.9 eV. The $f$-$p$ band exhibits nearly parabolic dispersion crossing the Fermi level. Comparing with the ARPES data, the calculated band structures of the $(\sqrt{3} \times \sqrt{3})R30°$ Eu-MLG superlattice is in general agree with experimental results structurally. The small discrepancies of the pocket shape around $K_F$ and the energy of 4$f$ bands can be attributed to difficulty in accurately calculating the strong correlation effect of Eu orbitals.

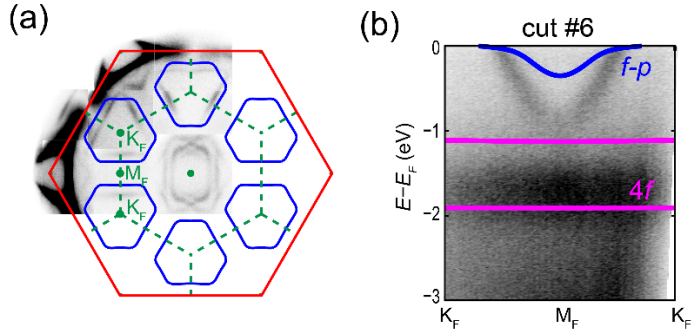

**Fig. S2. Calculated band structure of the $(\sqrt{3} \times \sqrt{3})R30°$ Eu-MLG superlattice. (a)** Calculated Fermi surface mapping of the $(\sqrt{3} \times \sqrt{3})R30°$ Eu superlattice. The red hexagon is the BZ of MLG. The dashed green hexagon is the folded BZ of the $(\sqrt{3} \times \sqrt{3})R30°$ Eu-MLG superlattice. **(b)** Calculated band structure of the $(\sqrt{3} \times \sqrt{3})R30°$ Eu-MLG superlattice along the $K_F$-$M_F$-$K_F$ direction.

## C. Non-long-range magnetic order of Eu-MLG

We conducted the ex-situ magnetic measurement by superconducting quantum interference device vibrating sample magnetometer (SQUID-VSM, Quantum Design). The *M-H* loops shown in Figs. S3(a)&(b) indicate the paramagnetism of Eu-MLG at 5 K and 300 K. The field-cooling magnetization curve shown in Fig. S3(c) is smooth without any magnetic phase transition. These results indicate that there is no long-range magnetic order of Eu-MLG. Therefore, the intrinsic Dirac cone at the K/K' point show no splitting feature in ARPES [see the Fig. 2(i)&(j)].

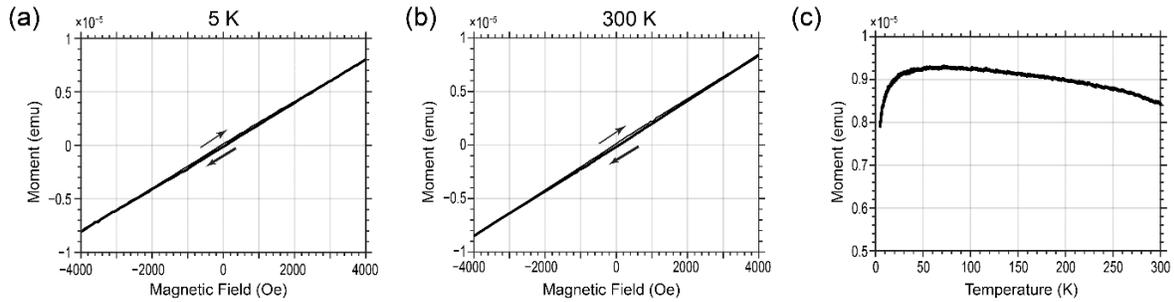

**Fig. S3. Magnetization measurements of Eu-MLG. (a)&(b)** *M-H* loops of Eu-MLG taken at 5 K (a) and 300 K (b), respectively. **(c)** Field-cooling magnetization curve of Eu-MLG with magnetic field of 4000 Oe.

## D. Tight-binding calculation based on the three-stage process

Figure S4(a) is the calculated tight-binding band structure of MLG that only involving the $\sqrt{3} \times \sqrt{3}$ superlattice BZ folding. Only the folded Dirac bands without splitting appear at the $\Gamma$ point. Further considering the Kekulé bond ordering, the folded Dirac bands will open a gap at the Dirac point as shown in Fig. S4(b). While simultaneously involving the exchange coupling with local moments, the folded Dirac bands will show a Zeeman-type spin-split as shown in Fig. S4(c).

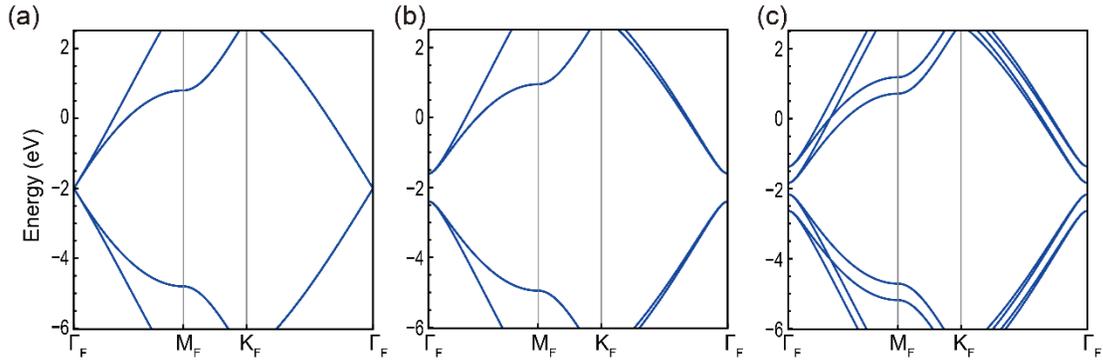

**Fig. S4. Tight-binding calculation results. (a-c)** Tight-binding calculation results based on only the $\sqrt{3} \times \sqrt{3}$ superlattice BZ folding (a), further including Kekulé bond ordering (b), and exchange coupling (c).